\documentclass[8.5pt,twoside,twocolumn]{article}
\usepackage[super,sort&compress,comma]{natbib} 
\usepackage{mhchem}
\usepackage{times,mathptmx}
\usepackage{sectsty}
\usepackage{balance} 

\usepackage{graphicx} 
\usepackage{lastpage}
\usepackage[format=plain,justification=raggedright,singlelinecheck=false,font=small,labelfont=bf,labelsep=space]{caption} 
\usepackage{fancyhdr}
\begin{document}

\thispagestyle{plain}
\fancypagestyle{plain}{
\renewcommand{\headrulewidth}{1pt}}
\renewcommand{\thefootnote}{\fnsymbol{footnote}}
\renewcommand\footnoterule{\vspace*{1pt}%
\hrule width 3.4in height 0.4pt \vspace*{5pt}} 
\setcounter{secnumdepth}{5}

\makeatletter 
\def\subsubsection{\@startsection{subsubsection}{3}{10pt}{-1.25ex plus -1ex minus -.1ex}{0ex plus 0ex}{\normalsize\bf}} 
\def\paragraph{\@startsection{paragraph}{4}{10pt}{-1.25ex plus -1ex minus -.1ex}{0ex plus 0ex}{\normalsize\textit}} 
\renewcommand\@biblabel[1]{#1}            
\renewcommand\@makefntext[1]%
{\noindent\makebox[0pt][r]{\@thefnmark\,}#1}
\makeatother 
\renewcommand{\figurename}{\small{Fig.}~}
\sectionfont{\large}
\subsectionfont{\normalsize} 

\fancyfoot{}
\fancyfoot[RO]{\footnotesize{\sffamily{1--\pageref{LastPage} ~\textbar  \hspace{2pt}\thepage}}}
\fancyfoot[LE]{\footnotesize{\sffamily{\thepage~\textbar\hspace{3.45cm} 1--\pageref{LastPage}}}}
\fancyhead{}
\renewcommand{\headrulewidth}{1pt} 
\renewcommand{\footrulewidth}{1pt}
\setlength{\arrayrulewidth}{1pt}
\setlength{\columnsep}{6.5mm}
\setlength\bibsep{1pt}

\twocolumn[
  \begin{@twocolumnfalse}
\noindent\LARGE{\textbf{Internal state thermometry of cold trapped
    molecular anions}}
\vspace{0.6cm}

\noindent\large{\textbf{Rico Otto,\textit{$^{a,b,\ddag}$},
    Alexander von Zastrow\textit{$^{b,\ddag}$}, Thorsten Best,\textit{$^a$} and
    Roland Wester\textit{$^{a,\ast}$}}}\vspace{0.5cm}

\noindent\textit{\small{\textbf{Received Xth XXXXXXXXXX 20XX, Accepted Xth XXXXXXXXX 20XX\newline
First published on the web Xth XXXXXXXXXX 200X}}}

\noindent \textbf{\small{DOI: 10.1039/b000000x}}
\vspace{0.6cm}

\noindent \normalsize{Photodetachment spectroscopy of OH$^-$ and
  H$_3$O$_2^-$ anions has been performed in a cryogenic 22-pole
  radiofrequency multipole trap. Measurements of the detachment cross
  section as a function of laser frequency near threshold have been
  analysed. Using this bound-free spectroscopy approach we could
  demonstrate rotational and vibrational cooling of the trapped anions
  by the buffer gas in the multipole trap. Below 50\,K the OH$^-$
  rotational temperature shows deviations from the buffer gas
  temperature, and possible causes for this are discussed. For
  H$_3$O$_2^-$ vibrational cooling of the lowest vibrational quantum
  states into the vibrational ground state is observed. Its
  photodetachment cross section near threshold is modelled with a
  Franck-Condon model, with a detachment threshold that is lower, but
  still in agreement with the expected threshold for this system.  
}
\vspace{0.5cm}
 \end{@twocolumnfalse}
  ]


\footnotetext{\textit{$^{a}$~Institut f\"ur Ionenphysik und Angewandte
    Physik, Universit\"at Innsbruck, Technikerstra\ss e 25, A-6020
    Innsbruck, Austria. }}

\footnotetext{\textit{$^{b}$~Physikalisches Institut,
    Albert-Ludwigs-Universit\"at Freiburg, Hermann-Herder-Str. 3,
    79104 Freiburg, Germany. }}


\footnotetext{\ddag~Both authors contributed equally to this
  work. Present addresses: R.~O.~ is now at the Department of
  Chemistry and Biochemistry, University of California San Diego, USA,
  and A.~v.~Z.~ is now at the Institute for Molecules and Materials,
  Radboud University Nijmegen, The Netherlands}

\section{Introduction}

Buffer-gas cooling of trapped molecular ions has been established as a
standard technique for producing cold molecules in recent years
\cite{asvany2009:ijm,wester2009:jpb}. The method is efficient both for
translational and for internal degrees of freedom, but precise
temperature measurements have only been performed for a limited number
of molecular ions. Most temperature measurements rely on high
resolution spectroscopy that resolves the Doppler profile to determine
the kinetic temperature of the trapped ions
\cite{schlemmer1999laser,mikosch2004:jcp,Glosik06,asvany2007:jcp,kreckel2008:jcp,asvany2008:prl}. Effective
temperatures for the internal degrees of freedom have also been
determined using rotationally-resolved bound-bound spectroscopy for
rotational temperatures
\cite{schlemmer1999laser,schlemmer2002laser,dzhonson2006:jms} or, more
rarely, using a vibrational hot-band analysis for the vibrational
temperature \cite{stearns07}. All of these techniques rely on
bound-bound action spectroscopy and are thus not applicable when
efficient probing schemes are not available, which is often the case
for molecular anions.

For negative molecular ions, however, photodetachment of the excess
electron provides a means of action spectroscopy that is applicable to
essentially all negative ions. Although at first thought it may seem
surprising that the non-resonant photodetachment process, by which an
electron may be lifted into the continuum as soon as the photon energy
exceeds the electron affinity of the resulting neutral molecule,
actually provides internal-state information on the initial anionic
molecule.  However, one should keep in mind that while the detached
electron may have any amount of kinetic energy, the molecular internal
structure typically prevails, so that well-defined propensity rules
exist for the discrete set of possible associated state-to-state
transitions. The total cross-section for photodetachment at any given
photon energy is thus made up of the sum of all open channels at that
energy, weighted by the associated transition strengths and initial
state populations. Therefore, by measuring the energy-differential
cross section over a region where different channels open, we can
determine the population of the corresponding initial states of the
molecular anion.

In this paper we demonstrate near-threshold bound-free photodetachment
spectroscopy as a thermometry scheme for molecular anions that are
trapped in a multipole radio-frequency ion trap. We present results
for the hydroxyl anion OH$^-$, one of the simplest molecular anions,
and for the hydroxyl-water cluster anion H$_3$O$_2^-$, a prototype for
anion-molecule clusters. The ions are subjected to buffer-gas cooling
in cold helium gas, the temperature of which has been varied between
ten Kelvin and room temperature.

\section{Experimental Methods}

The setup employed in these experiments has been described in parts
previously \cite{best2011:apj}. In brief, molecular anions are
produced in an electrical discharge in a pulsed gas jet, and extracted
using a Wiley-McLaren type mass spectrometer. The time-of-flight
selected ions are then stored in a 22-pole radio-frequency ion trap
\cite{wester2009:jpb}. The trap is mounted on a closed-cycle
refrigerator and is equipped with ohmic heaters and a thermal
radiation shielding. This allows to fix the trap temperature anywhere
between 8\,K and room temperature, measured with a silicon diode
thermometer. {\bf Trapped ions are thermalized by collisions with an
  inert buffer gas (helium) that is applied into the trap enclosure at
  typical densities of 10$^{14}$cm$^{-3}$ leading to 10$^4$ - 10$^5$
  collisions per second.}

After an initial thermalisation time {\bf of 10 ms}, the anions are
exposed to a laser beam propagating along the trap axis for the
selected exposure time, after which the remaining ions are detected
using the same time-of-flight mass spectrometer. By varying the
exposure time, we measure the laser-induced ion loss rate, which is
directly proportional to the energy-differential photodetachment cross
section at the given photon energy. The full energy-differential cross
section is determined by tuning the laser wavelength, which is
referenced to a high-resolution wavemeter. The laser light is derived
from external cavity red diode lasers for photodetachment of OH$^-$
anions, and free-running bluray diode lasers for H$_3$O$_2^-$, which
have been built in-house. In either case, the lasers can be tuned over
several nanometres using temperature tuning. By coupling the laser
light through a single-mode optical fibre temperature-induced drifts
of the beam pointing are suppressed. Therefore, the laser always
illuminates the same spot within the ion cloud, and a full tomographic
measurement \cite{trippel2006:prl,hlavenka2009:jcp,best2011:apj} at
every laser wavelength is not required for the relative cross section
measurements.

\begin{figure}[tb]
\center
\includegraphics[width=\columnwidth]{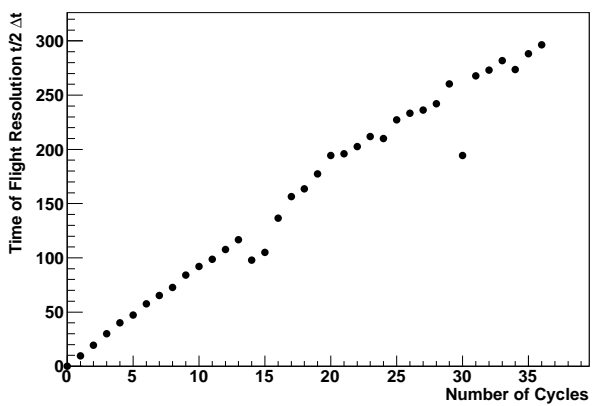}
\caption{Measured time resolution of OH$^-$ ions after a selected
  number of round trips in the multicycle reflectron mass
  spectrometer. Plotted is half of the time-of-flight resolution,
  which directly represents the mass resolution achieved with the
  device.}
\label{FIG7}
\end{figure}

The ions that remain in the trap after being exposed to the
photodetachment laser, are extracted from the trap and detected. In
order to distinguish different ionic species the ions enter a
multicycle reflectron time-of-flight mass spectrometer. This device
consists of two coaxial reflectron mirrors that are placed 460\,mm
apart, similar to a resonator ion trap
\cite{zajfman1997:pra,dahan1998:rsi}. While the rear mirror is held on
a repulsive potential the front mirror is grounded as the ion packet
enters the device. As this mirror is switched to a repulsive potential
as well, the ion packet becomes confined in the multicycle reflectron
and travels back and forth between the two mirrors at a fixed mass
dependent frequency. Since in a time-of-flight arrangement the mass
resolution m/$\Delta$m = t/2$\Delta$t is directly depending on the
flight time, the multicycle reflectron design offers an elegant way to
achieve a desired mass resolution by confining the ions for a given
number of cycles. Fig.\ \ref{FIG7} displays the mass resolution for
OH$^-$ achieved using the multicycle reflectron as a function of the
first 35 cycles. When the desired mass resolution is achieved the ion
packet is released by switching the front mirror to ground again. The
ions leave the device and get detected on a multi channel plate ion
detector.

\section{Results and Discussion}

\subsection{Near-threshold photodetachment of OH$^-$}
\label{oh:sect}

For the hydroxyl molecular anion OH$^-$ in the radiofrequency 22-pole
ion trap, the vibrational motion can be considered to be frozen out
completely (i.\ e.\ there is on average far less than one
excited-state ion present in the trap) already at room temperature. At
low temperatures we therefore only have to deal with rotational
excitations of an approximately rigid rotor with a rotational constant
of $B_0=18.735$\,cm$^{-1}$ \cite{matsushima2006:jms}. At 50\,K,
rotational levels up to $J=2$ are populated to a measurable
degree. Measurements have therefore been performed for a range of
photon energies covering the thresholds of the lowest three rotational
states, as described in the experimental methods section below.

\begin{figure}[tb]
\center \includegraphics[width=\columnwidth]{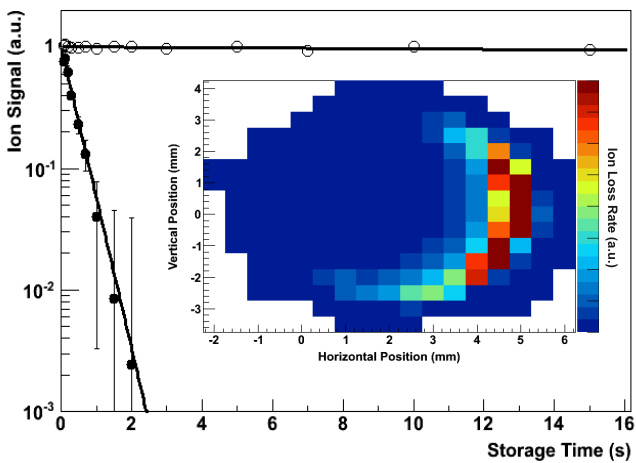}
\caption{Laser-induced trap loss of OH$^-$ above the $J=0$
  photodetachment threshold. All rotational states contribute to this
  decay channel. The inset shows a two-dimensional tomography scan,
  which represents the column density of the ions in the 22-pole trap
  (see methods).}

\label{FIG1}
\end{figure}

The threshold for photodetachment of ground state ($J=0$) hydroxyl
anions lies at 14740.982(7)\,cm$^{-1}$, the electron affinity of
neutral $^{16}$O$^1$H \cite{goldfarb2005:jcp}. When tuning the
photodetachment laser to a slightly larger photon energy,
photodetachment dominates the ion loss from the trap, as shown in
Fig.\ \ref{FIG1}. The long background lifetime of the ions in the trap
of more than 10$^3$\,s allows for photodetachment rate measurements
over several orders of magnitude. The inset in Fig.\ \ref{FIG1} shows
a two-dimensional tomography scan of the spatial density of the
trapped ion cloud, which is used in the determination of absolute
photodetachment cross sections
\cite{trippel2006:prl,hlavenka2009:jcp}.

\begin{figure}[tb]
\center \includegraphics[width=\columnwidth]{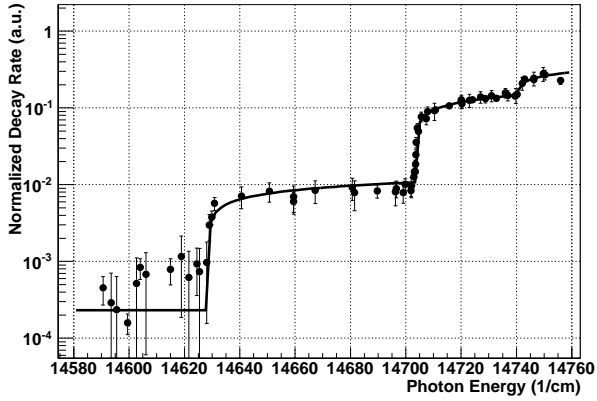}
\caption{Photodetachment cross section of OH$^-$ at a trap temperature
  of 50\,K for varying photon energy. The steps in the cross section
  are due to the opening of loss channels corresponding to the $J=2,1$
  and $0$ rotational states of the anion.}
\label{FIG2}
\end{figure}

For photon energies below the $J=0$ detachment threshold, only
rotationally excited hydroxyl anions can be detached. The measured
relative rate of photodetachment of the trapped ions is plotted in
Fig.\ \ref{FIG2} as a function of the photon energy, which is tuned
across the detachment thresholds for the first three rotational
states. The scan was obtained for a helium buffer gas temperature of
50\,K. The thresholds for the states with $J=0$, 1, and 2 are clearly
observed. They are assigned to the openings of the photodetachment
transitions R3(0), Q3(1), and P3(2) (for the notation see
Ref.\ \cite{goldfarb2005:jcp}) transitions, which all form neutral OH
in the rotational ground state of the $^2\Pi_{3/2}$ spin-orbit
manifold. Thus, the energy difference between the thresholds
represents directly the rotational levels in the molecular anion,
which are known from rotational spectroscopy \cite{matsushima2006:jms}
to be given by 37.5\,cm$^{-1}$ for the $J=1 \leftarrow 0$ excitation
and 112.3\,cm$^{-1}$ for the $J=2 \leftarrow 0$ excitation,
respectively. The measured differences of the thresholds agree very
well with this prediction.

The shape of the cross section near threshold, more specifically the
relative heights of the components that terminate at the different
thresholds, is determined by the population of the different
rotational states of the anion. To describe the shape of the cross
section we employ the functional form
\begin{equation}
  \sigma(h \nu) \propto \sum_J P(J) · I_{J}
  \left(h \nu - \epsilon_{J}\right)^p,
\label{XSECFIT}
\end{equation}
where $h \nu$ is the photon energy, $P(J)$ is the population of the
rotational levels of the hydroxyl anion, $I_J$ and $\epsilon_J$ are
the corresponding H{\"o}nl-London factors (taken from
Ref.\ \cite{goldfarb2005:jcp}) and threshold energies, and $p$ is the exponent
for the power law describing the energy dependence of the cross
section near threshold. For a non-interacting outgoing s-wave electron
$p=0.5$, corresponding to the Wigner threshold law. In the present
case the electron-dipole coupling modifies the power law
\cite{smith1997:pra}. For this scenario the exponent has been derived to be
$p=0.28$ \cite{ENGELKING1982}.

By fitting Eq.\ \ref{XSECFIT} to the measured cross section in
Fig.\ \ref{FIG2}, we obtain the population in the rotational levels
$J=0\dots 2$. When leaving $p$ as free parameters in the fit, we
obtain $p_P=0.23(4)$ and $p_Q=0.30(2)$ for the P- and Q-branches,
respectively. As these values are in good agreement with the value
derived in Ref.\ \cite{ENGELKING1982}, we fix the parameter to the
latter value $p=0.28$ to improve the stability of the fit. This leaves
the rotational populations as the only free fit parameters. From the
population ratio of two rotational states the rotational temperature
can be derived. The most precise evaluation is based on the ratio of
molecules in $J=2$ and $J=1$, since these two thresholds have been
covered best in the frequency scan. This yields a rotational
temperature of $57\pm6$\,K for the data in Fig.\ \ref{FIG2}, in
reasonable agreement with the helium buffer gas temperature of 50\,K.

\begin{figure}[tb]
\center
\includegraphics[width=\columnwidth]{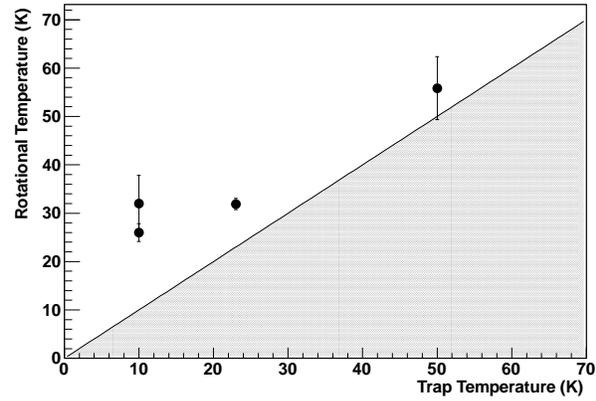}
\caption{Rotational temperatures of OH$^-$ anions determined from the
  ratio of the observed photodetachment transitions originating in
  $J=1$ and $J=2$ levels, respectively, for different temperatures of
  the helium buffer gas in the 22-pole trap.}
\label{FIG3}
\end{figure}

Further spectroscopic scans of the photodetachment rate near threshold
have been performed for buffer gas temperatures of 10 and 23\,K and
the spectra have also been fit with
Eq.\ \ref{XSECFIT}. Fig.\ \ref{FIG3} compares the obtained rotational
temperatures to the buffer gas temperatures in the corresponding
measurements. While at higher temperatures, the temperatures agree
roughly within the experimental uncertainty, a significant deviation
is found for lower temperatures, where the ion temperature always
exceeds the buffer gas temperature.

Different causes may be responsible for this deviation. Radiofrequency
heating due to collisions with buffer gas atoms in the presence of the
radiofrequency field is known to increase the translational
temperature of trapped ions. This will indirectly also affect the
rotational temperature. However, this effect is suppressed in traps
with a high multipole order such as the employed 22-pole trap to the
level of a few percent \cite{asvany2009:ijm,wester2009:jpb}. It may,
however, play a larger role if the effective potential is disturbed by
surface patch potentials or by pockets that are induced by slight
geometrical distortions of the trap electrodes
\cite{otto2009:jpb}. The two different rotational temperatures
measured for 10\,K buffer gas temperature in Fig.\ \ref{FIG3}, even
though barely statistically significant, may be caused by different
patch potential conditions in the ion trap. A second possible
mechanisms is the direct rotational excitation by room temperature
blackbody radiation that enters the trap via its entrance and exit
electrodes. Incomplete thermalisation can be excluded, because many
hundred to a few thousand buffer gas collisions have occured even for
the shortest storage times that we employed in the measurements. In
the future, we will carry out further investigations by varying the
trapping potentials and the buffer gas conditions in the trap, in
order to clarify the source of the observed rotational heating.

\subsection{Near threshold photodetachment of H$_3$O$_2^-$}

{\bf For small molecular systems where the rotational structure can be
  clearly resolved the above method should be of general use. In order
  to test the applicability for larger molecules and probe its
  sensitivity on vibrational cooling the photodetachment thermometry
  method is now applied to a more complex molecular system.}

H$_3$O$_2^-$ is a model system for a molecular cluster that may rotate
and vibrate even at low temperatures, because of several low-frequency
vibrational modes. It represents the smallest deprotonated water
cluster, small enough that full nine-dimensional Born-Oppenheimer
potential calculation and detailed quantum calculations of the
fundamental vibrational frequencies have been carried out
\cite{huang2004:jacs,mccoy2005:jcp}. It has been found that the ground
state structure of H$_3$O$_2^-$ is an inversion-symmetric bent
configuration with the central proton shared between hydroxyl
anions. As a consequence, the low-frequency vibrational modes are
split due to tunnelling.

\begin{figure}[tb]
\center
\includegraphics[width=\columnwidth]{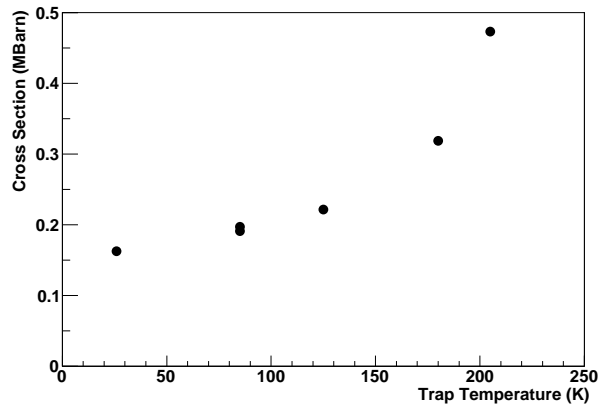}
\caption{Photodetachment cross section of H$_3$O$_2^-$ at a photon
  energy of $24410 \pm 20\,\textrm{cm}^{-1}$ for different trap
  temperatures (1\,MBarn = 10$^{-18}$\,cm$^2$) An independent full 2D
  tomographic measurement has been performed at each temperature.}
\label{FIG4}
\end{figure}

The threshold for ground state H$_3$O$_2^-$ photodetachment lies near
3.0\,eV or about 24200\,cm$^{-1}$, based on the electron affinity of
OH of 1.82\,eV and the dissociation energy of H$_3$O$_2^-$ of 1.18\,eV
\cite{arnold1995:jcp} and neglecting possible bound states in the
neutral OH-water complex. We have therefore measured the
near-threshold photodetachment cross sections at a fixed photon energy
of $24410 \pm 20$\,cm$^{−1}$ for trap temperatures of 30, 85, 130, 180
and 205\,K respectively, using the tomographic procedure outlined in
the methods section. The result is shown in Fig.\ \ref{FIG4}. At room
temperature, where we could only complete a partial tomography due to
the increase in width of the density distribution, we estimate a cross
section of $1.5 \pm 0.5$\,MBarn (1\,MBarn = 10$^{-18}$\,cm$^2$) by
extrapolation. The data in Fig.\ \ref{FIG4} show a strong increase of
the cross section with temperature, which also requires a strong
dependence of the detachment cross section on the internal quantum
state. In turn, the change in cross section also shows that cooling of
these internal quantum states actually takes place when lowering the
buffer gas temperature from room temperature down to 30\,K.

\begin{figure}[tttb]
\center
\includegraphics[width=0.9\columnwidth]{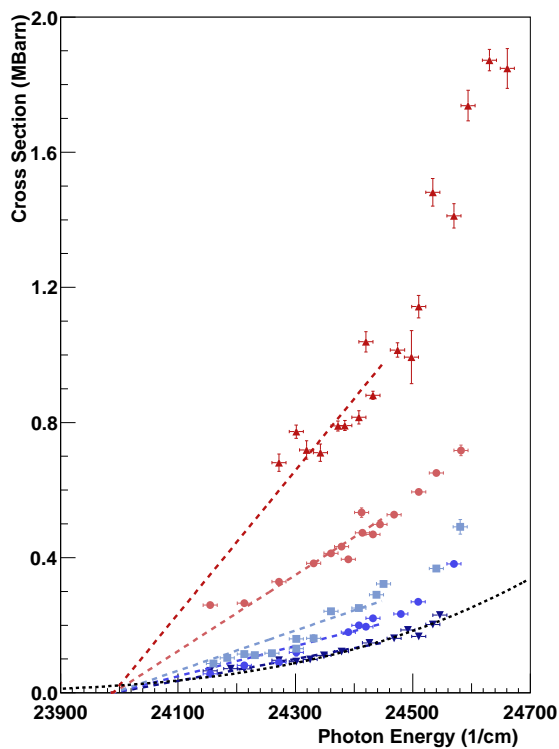}
\caption{Near threshold photodetachment scans of H$_3$O$_2^-$ for
  temperatures of 30, 85, 130, 205 and 300\,K, respectively (from
  bottom to top and from dark blue to dark red).  Each measurement is
  based on a partial tomography of the cloud density, scaled according
  to a single full tomography at each temperature. The straight dashed
  lines represent simple linear extrapolations of the cross sections
  to identify the apparent photodetachment threshold. The black dotted
  line is the result of our Franck-Condon model for the detachment of
  ground state H$_3$O$_2^-$ and agrees well with the data for 30\,K.}
\label{FIG5}
\end{figure}

To gain further insight, the photodetachment cross section is measured
as a function of the laser frequency in the vicinity of the detachment
threshold. The results are plotted in Fig.\ \ref{FIG5} for
temperatures of 30, 85, 130, 205 and 300\,K, respectively. For all
temperatures the cross section increases with a nearly linear photon
energy dependence and with an apparent threshold of about
24000\,cm$^{-1}$. The tuning range of the employed diode laser was
limited to energies above 24150\,cm$^{-1}$, which prevented
measurements closer to this apparent threshold. Nevertheless it
becomes evident that the energy dependence of the cross section is
quite different from that for bare OH$^-$. Describing it with a
function proportional to $(h \nu- E_0)^p$ near an energy threshold
$E_0$, the data suggests $p$ to be approximately 1 for H$_3$O$_2^-$
and not about $0.28$ as found for OH$^-$ or 0.5 for pure s-wave
detachment. This can not be easily explained by the electronic
structure of the H$_3$O$_2^-$ complex, and is more likely caused by
its internal structure, as discussed below.

For all studied temperatures in Fig.\ \ref{FIG5} a linear
extrapolation fits the data well. The intercept of this extrapolation
with the photon energy axis is approximately the same for all
temperatures and its value of about 24000\,cm$^{-1}$ agrees well with
the expected threshold of 24200\,cm$^{-1}$, in particular when
considering that the latter value has an estimated accuracy of several
hundred wavenumbers. The slope, however, changes significantly, which
is again a signature of cooling of the internal temperature of the
trapped H$_3$O$_2^-$ anions at the different buffer gas
temperatures. Specifically, it can be caused by the changes of the
rotational or vibrational state population. The rotational states are
very closely spaced, given the rotational constants of H$_3$O$_2^-$ of
10, 0.3 and 0.3\,cm$^{-1}$ \cite{mccoy2005:jcp}. Therefore all
dipole-allowed detachment transitions are also energetically allowed
at the photon energies of the measurement. The H{\"o}nl-London factors
for all dipole-allowed photodetachment transitions of a specific
rotational state of the anion are assumed to sum up to unity, similar
to the case for OH$^-$ where this renders the absolute photodetachment
cross section independent of temperature \cite{hlavenka2009:jcp}. This
implies that a change of the rotational population due to lowering the
temperature will not have an influence on the measured cross
section. The observed change therefore has to be attributed to
vibrational cooling of the cluster anion and not to rotational
cooling.

The lowest vibrational frequencies of the cluster are two frequencies
$\omega_+, \omega_-$ of the torsional mode of about 132 and
215\,cm$^{-1}$, which are split due to tunnelling between different
equivalent structures on the Born-Oppenheimer potential surface
\cite{mccoy2005:jcp}. At 30\,K one expects from Boltzmann statistics
about 0.1\% in the lowest excited vibrational state. Even if the
vibrational temperature of the anions were several Kelvin larger than
the buffer gas temperature, which may occur due to trap imperfections
and radiofrequency heating (see also the discussion in section
\ref{oh:sect}) \cite{wester2009:jpb}, still no substantial population
in any of the excited state is expected. Then at 85\,K about 11\% of
the anions should be found with one of the two torsional modes
excited. Higher-frequency vibrational modes and also overtones of the
low frequencies only become populated at higher temperatures. At
200\,K many states are populated, but due to the lack of knowledge on
the eigenenergies of the overtone states we did not calculate
populations for this temperature.

For the two torsional eigenstates, which are symmetric and
antisymmetric linear combinations of two localised torsional modes, we
assume that their Franck-Condon factors upon photodissociation to the
neutral OH-H$_2$O system, which subsequently dissociates into OH and
H$_2$O, are the same. One can then use the measured detachment cross
sections for 30\,K and 85\,K, which increase by about a factor of 1.2
(see Fig.\ \ref{FIG4}), to extract that the 11\% torsionally excited
anions should have about a threefold enhanced detachment cross section
compared to vibrational ground state clusters.

\begin{figure}[tb]
\center
\includegraphics[width=\columnwidth]{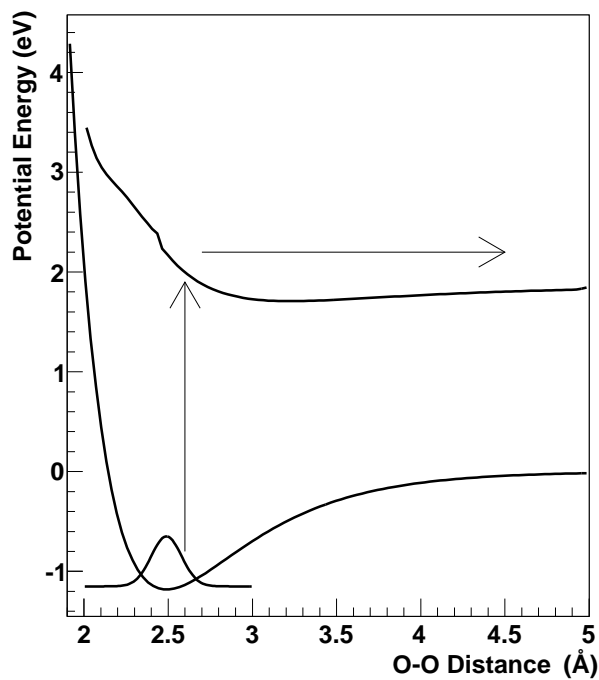}
\caption{Franck-Condon model for the photodetachment threshold
  behaviour based on the variation of potential energy along the O-O
  stretch coordinate. The Franck Condon overlap is determined by the
  ground state wavefunction at the Condon point corresponding to the
  incoming photon and released electron energy. The energy threshold
  for photodetachment into the dissociative continuum is determined
  from the H$_3$O$_2^-$ dissociation energy $E_\textrm{diss}$ and the
  hydroxyl electron affinity $EA$. The neutral potential curve was
  taken from an ab initio calculation.}
\label{FIG6}
\end{figure}

To understand the threshold behaviour of the cross section we have
developed a qualitative Franck Condon model. A full theoretical model,
which would be required in order to use the presented measurements for
calibration-independent vibrational thermometry, is beyond the scope
of this work. Here we restrict ourselves to a one-dimensional
description along the O-O direction and assume that upon
photodetachment the neutral molecule will dissociate primarily along
this direction. We fix all other coordinates to the equilibrium values
for the anion {\bf that were obtained from a MP2/aug-cc-PVDz geometry
  optimization} and assume that no significant geometry changes along
orthogonal directions occur upon photodetachment. {\bf Our calculation
  does not account for zero point energy which is a prerequisit to
  find the symmetric (HO$\cdots$H$\cdots$OH)$^-$ anion geometry.
  Instead the equilibrium structure used in this model is described by
  a OH$^-$(H$_2$O) cluster geometry.}  A sketch of the relevant
potential curves in this one-dimensional model is shown in
Fig.\ \ref{FIG6}. The anion ground state wavefunction is centred
around the equilibrium distance at 2.4896\,\AA. It can be approximated
by a Gaussian with an r.m.s. width calculated from the eigenfrequency
of the O-O stretch mode of 515\,cm$^{-1}$ \cite{mccoy2005:jcp}. The
energy difference between the anion ground state and the neutral
asymptote OH + H$_2$O is assumed to be 3.0\,eV or 24200\,cm$^{-1}$
\cite{arnold1995:jcp}. To obtain the potential curve for the
dissociating neutral complex, we have performed one-dimensional ab
initio potential energy calculations at the CCSD(T) level of theory in
the aug-cc-PVDz basis. Neglecting zero-point energies, this potential
is also shown in Fig.\ \ref{FIG6}, it features a shallow minimum
around $\approx$ 3.2\,\AA. In the range between 2.1\,\AA and 2.7\,\AA,
the ab initio potential is repulsive and lies higher in energy than
the asymptote.

The photodetachment cross section depends on the Franck-Condon overlap
between the anion and neutral vibrational wavefunctions and on the
amount of excess energy that is released to the free electron. As the
Franck-Condon overlap with the bound neutral states residing in the
shallow potential well is small, detachment into bound states is
expected to be negligible. The neutral wavefunctions in the
dissociation continuum are simplified by delta functions at the inner
turning points of the potential. The Franck-Condon factor then becomes
the square of the anion vibrational wavefunction at a fixed O-O
distance. The excess energy taken by the electron is given by the
difference between the photon energy, the threshold energy for
photodetachment and the repulsive energy between the neutral
fragments. As the photon energy is increased, larger repulsive
energies can be reached, so that smaller O-O distances become
accessible for photodetachment. Due to the Franck Condon factor this
enhances the cross section much more than it is reduced by the
decreasing electron excess energy. Qualitatively, this causes the
approximately linear increase of the cross sections in
Fig.\ \ref{FIG5}.

More specifically, the photodetachment cross section can be obtained
by integrating over all contributing O-O distances. Using a similar
threshold scaling for the electron kinetic energy as for OH$^⁻$ in the
previous section (see Eq.\ \ref{XSECFIT}), the cross section here is
given by
\begin{equation}
\sigma(h \nu)
\propto 
\int_{(h \nu - E_0 - V(r)) > 0}
|\psi(r)|^2
(h \nu - E_0 - V(r))^p
\;dr,
\label{EQ:CROSSSECTION}
\end{equation}
where $h \nu$ is again the photon energy, $E_0$ is the energy
threshold of the cluster, $V(r)$ is the repulsive potential of the
neutral complex, and $\psi(r)$ is the anion wave function as a
function of the O-O distance $r$. For the exponent $p$ we assume 0.5
for s-wave electron detachment, which should be a better assumption
here than for OH$^-$ due to the vanishing electric dipole moment of
the inversion symmetric ground state of H$_3$O$_2^-$. The result of
this model is overlaid with the 30\,K data in Fig.\ \ref{FIG5}, where
both a reduction of the energy threshold to 23700\,cm$^{-1}$ and a
scaling factor had to be applied to fit to the experimental
data. Qualitatively, the model reproduces the shape of the 30\,K data,
identified as photodetachment from the vibrational ground state of the
anion. Also the asymptotic cross section for photon energies high
above threshold where the Franck-Condon factor approaches unity falls
into the same order of magnitude of $10^{-17}$\,cm$^2$ as for
OH$^-$. The lowering of the energy threshold by 500\,cm$^{-1}$ can
either be caused by the finite accuracy of the neutral potential
calculation or by the uncertainty of the dissociation energy of the
anion. In future work it would be desirable to independently obtain
either $E_0$ or $V(r)$ with higher accuracy, because then the
detachment data can be used to put stronger constraints on either the
dissociation energy of the H$_3$O$_2^-$ cluster or the dissociative
potential of the OH + H$_2$O collision system.

Upon increasing the temperature to 85\,K, the tunnelling doublet of
the torsional mode becomes excited. To first approximation, this
leaves the shape of the ground state wavefunction of the O-O-stretch
mode unaffected. In this situation, a $\Delta\,\nu_{\rm torsion} = -1$
photodetachment transition becomes possible, which allows for a larger
electron kinetic energy. Although the Franck-Condon overlap is
supposedly small for this type of transition, the extra electron
energy may increase the total cross section. The one-dimensional model
described above can not treat this case. A quantitative evaluation of
the change in cross section for the torsionally excited vibrational
states already requires a multi-dimensional calculation beyond the
scope of this work. As temperatures increase beyond 100\,K, overtones
of the torsional mode become excited. At this point, even qualitative
predictions are difficult, as the corresponding energies and
wavefunctions are largely unknown. Furthermore, it seems conceivable
that also rotational excitation may affect the Franck Condon overlap
and thereby the cross section by means of centrifugal stretching of
the O-H-O bonds.

\section{Conclusion}

We have shown that near-threshold photodetachment spectroscopy of
trapped and buffer gas cooled molecular anions is capable of measuring
internal state populations. This will allow further investigations of
the efficiency of thermalisation of the rotational and vibrational
degrees of freedom by buffer gas cooling under different trapping
conditions. The information is also useful for internal
state-dependent collision and reaction studies, and has recently
already been employed in reactions of H$_3$O$_2^-$ with CH$_3$I
\cite{otto2012:natc,otto2012:fd}. Furthermore, photodetachment
spectroscopy is an important prerequisite for high resolution
terahertz spectroscopy of low-lying rotational and vibrational states
in such systems, because it allows for action spectroscopic detection
of inernal excitations on the few-molecule level. While our method is
in principle quite general and can be applied to essentially any
anionic species, a sufficient understanding of the molecular structure
of both the anion and the corresponding neutral is necessary.

\section{Acknowledgements}

R.O. acknowledges support by the Landesgraduiertenf{\"o}rderung
Baden-W{\"u}rttemberg. This work is supported by the European Research
Council under ERC grant agreement No. 279898. We thank Matthias
Weidem{\"u}ller for many stimulating discussions and Petr Hlavenka and
Sebastien J{\'e}zouin for their help during preliminary
experiments. We also thank the University of Freiburg, where the
measurements presented here have been carried out, for supporting this
research.


\balance


\footnotesize{

\providecommand*{\mcitethebibliography}{\thebibliography}
\csname @ifundefined\endcsname{endmcitethebibliography}
{\let\endmcitethebibliography\endthebibliography}{}

}

\end{document}